\documentclass[a4paper]{PoS}

\usepackage{amssymb}
\usepackage{amsmath}
\usepackage{url}
\usepackage{rotating}
\usepackage{float}
\usepackage{siunitx}
\usepackage{graphicx}
\newcommand{\ratio}{{$\sigma_{e^+p}/\sigma_{e^-p}$} }

\newcommand{\ep}{{$e^- p$} }
\newcommand{\pp}{{$e^+ p$} }
\newcommand{\pmp}{{$e^\pm p$} }
\newcommand{\rtg}{{$R_{2\gamma}$} }

\makeatletter
\g@addto@macro\bfseries{\boldmath}
\makeatother

\title{Results from the OLYMPUS Experiment on the Contribution of Hard Two-Photon Exchange to Elastic Electron-Proton Scattering}

\ShortTitle{Results from the OLYMPUS Experiment on Two-Photon Exchange in \ep Scattering}

\author{\speaker{Brian S. Henderson}\thanks{For the OLYMPUS Collaboration.}\\
        Massachusetts Institute of Technology\\
        E-mail: \email{bhender1@mit.edu}}

\abstract{Measurements of the ratio of the elastic form factors of the proton ($\mu_pG_E/G_M$) exhibit
a strong discrepancy. Experiments using unpolarized beams and Rosenbluth separation
to determine the form factors have found values of the ratio approximately consistent
with unity over a wide range of $Q^2$, while polarization transfer experiments suggest that the ratio
decreases as a function of $Q^2$.  The most widely-accepted hypothesis to explain this discrepancy is that
hard two-photon exchange (TPE) significantly contributes to the elastic $ep$ cross section.  Hard TPE
has been neglected in previous analyses of electron-proton scattering scattering experiments,
in part due to the fact that there exists no model independent way to calculate the contribution.  The 
effect of hard TPE may be measured experimentally, however, via precise determination of the ratio of the electron-proton
and positron-proton elastic cross sections.  The OLYMPUS experiment collected more than 3 fb$^{-1}$ of exclusive \ep and \pp
elastic scattering data at DESY in 2012, and has determined the elastic \ratio ratio to unprecedented precision up to $Q^2\approx2.2$
(GeV/$c$)$^2$, $\epsilon\approx0.4$.  This presentation will discuss the OLYMPUS experiment and analysis, and present the recently
published results from OLYMPUS in the context of the results from the other two TPE experiments.}

\FullConference{XVII International Conference on Hadron Spectroscopy and Structure - Hadron2017\\
		25-29 September, 2017\\
		University of Salamanca, Salamanca, Spain}

\begin{document}

\section{Introduction}

\begin{figure}[thb!]
    \centering
    \includegraphics[width=0.55\columnwidth]{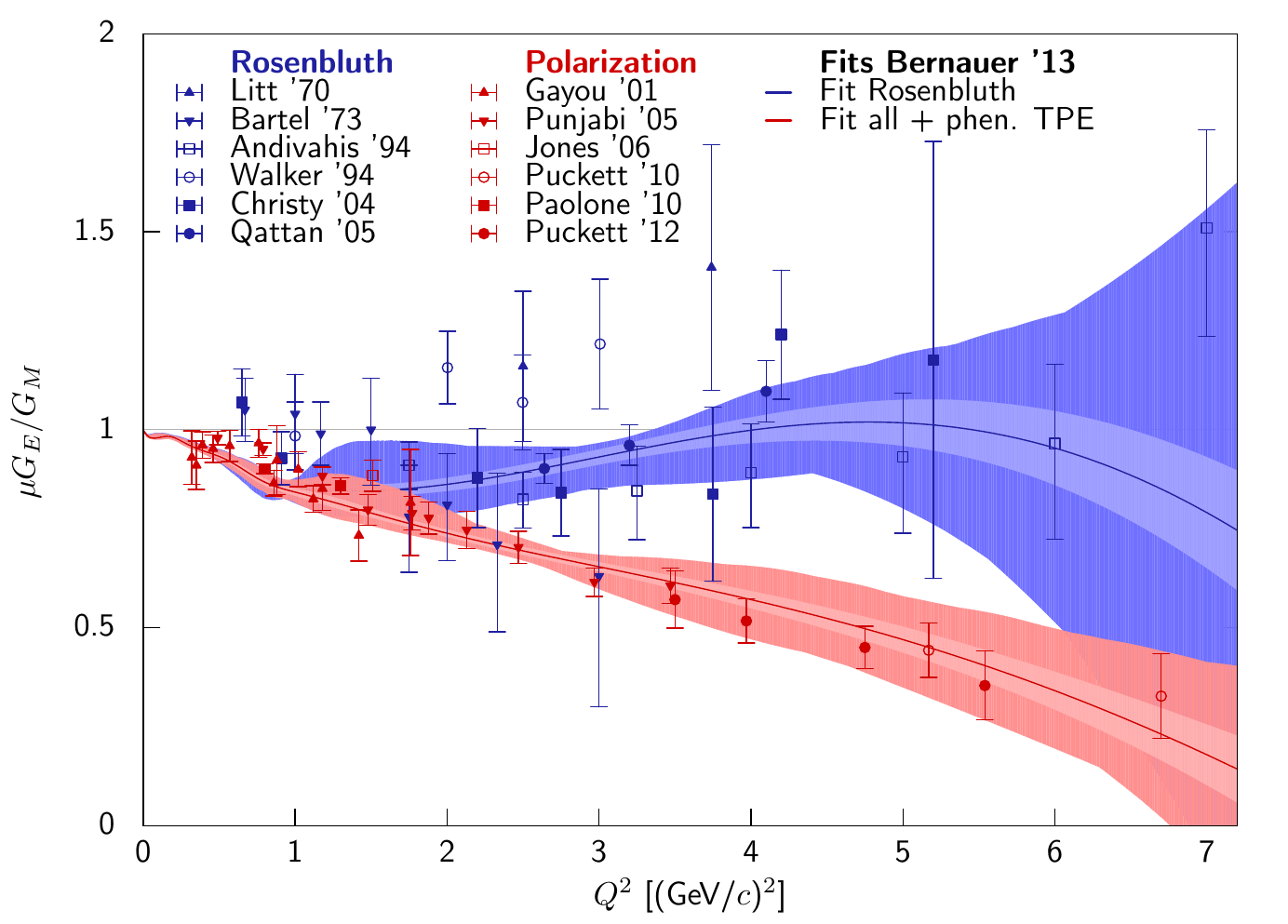}
    \caption{Selected measurements of the form factor ratio $\mu_pG_E(Q^2)/G_M(Q^2)$, illustrating the discrepancy between polarized \cite{pol11,pol3,pol12,pol7,pol9,pol13} and unpolarized
    (Rosenbluth separation) \cite{ff3,ff4,ff8,ff9,ff10,ff11} measurements, along with 
    phenomenological fits to the two data types \cite{BerFFPhysRevC.90.015206}. (Figure reproduced from Reference \cite{Milner:2014}.)}
    \label{fig:disc}
\end{figure}

The development of polarized electron beams and proton targets made possible new approaches to measuring the
elastic form factors of the proton in the 1990s \cite{pol11,pol3,pol12,pol7,pol9,pol13}.  While these
methods do not provide access to the individual form factors, $G_E(Q^2)$ and $G_M(Q^2)$, they do provide
a way of precisely measuring the ratio $\mu_pG_E(Q^2)/G_M(Q^2)$ in which many systematic uncertainties, such
as radiative corrections and absolute normalization, at least partially cancel.   Previous measurements of the proton form factors using
the Rosenbluth separation technique \cite{PhysRev.79.615} with inclusive $e^-p$ elastic scattering data
favored $\mu_pG_E(Q^2)/G_M(Q^2)$ consistent with unity up to $Q^2\sim 10$ (GeV/$c$)$^2$ \cite{ff3,ff4,ff8,ff9}.  As shown in Figure
\ref{fig:disc}, however, the measurements using the new polarization-based techniques showed a decreasing value
of the form factor ratio as a function of $Q^2$.  More modern Rosenbluth separation measurements using exclusive
event reconstruction \cite{ff10,ff11} and re-analysis of the previous data \cite{PhysRevC.68.034325} failed
to resolve the discrepancy.  Given that the form factors represent fundamental properties of nucleons, these
results precipitated renewed theoretical and experimental efforts to study elastic \ep scattering.

\begin{figure}[thb!]
    \centering
    \includegraphics[width=0.5\columnwidth]{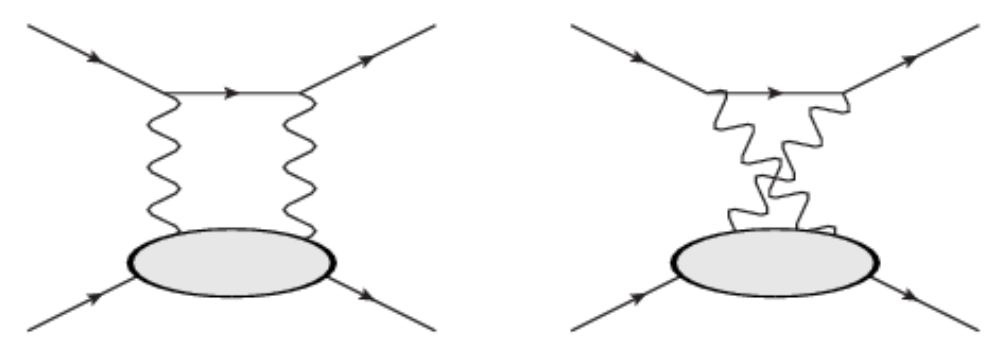}
    \caption{The two photon exchange diagrams (``box'' and ``crossed-box'') for elastic $e^-p$ scattering.  Since each photon carries significant four-momentum, the proton
             propagator between the vertices need not be that of an on-shell proton, making calculations of the contributions from these diagrams challenging.}
    \label{fig:2gamma}
\end{figure}

The predominant hypothesis that emerged for the explanation of the form factor ratio discrepancy was that hard two
photon exchange (TPE) effects, i.e., contributions from the Feynman diagrams shown in Figure \ref{fig:2gamma}, were improperly
neglected in previous Rosenbluth separation analyses \cite{Guichon:2003qm,Blunden:2003sp,Blunden:2005ew}.  While radiative
corrections, including ``soft'' TPE, are critical to form factor extractions in Rosenbluth separation experiments and several
standard prescriptions exist \cite{MoRevModPhys.41.205,MaximonPhysRevC.62.054320}, these prescriptions neglect contributions
from hard TPE.  Polarization-based experiments, in measuring a ratio of ratios of final state asymmetries, are relatively
insensitive to TPE contributions.  Due in part to the uncertainty in the proton's propagator between the two photon interactions, calculations of
the hard TPE contribution are model dependent and vary significantly in their effect on the form factor ratio.  While some
calculations suggest that hard TPE contributions are large enough to explain the entirety of the form factor discrepancy \cite{Afanasev:2005mp,Blunden:2005ew,PhysRevLett.103.092004},
others suggest that such effects should be negligible \cite{TomasiGustafsson:2009pw,PhysRevC.78.015205}.

Given the substantial uncertainty surrounding theoretical calculations of hard TPE, experimental measurements were needed to test
the hypothesis that hard TPE can explain the form factor ratio discrepancy.  An experimental signature of TPE may be elicited by
comparing elastic \ep and \pp scattering.  The interference of the one- and two-photon exchange diagrams (Figure \ref{fig:2gamma})
produces a term in the \pmp cross section that is odd in the sign of the lepton.  If other lepton-sign odd radiative effects, such as contributions
from soft bremsstrahlung are accounted for, then the matrix element corresponding to the TPE diagrams $\mathcal{M}_{\gamma\gamma}$ may be extracted from
the ratio \rtg of the elastic \pp and \ep cross sections:
\begin{equation}
 R_{2\gamma}\left(\epsilon,Q^2\right) = \frac{\sigma_{e^+p}\left(\epsilon,Q^2\right)}{\sigma_{e^-p}\left(\epsilon,Q^2\right)} = 1 + \frac{4\Re\left[ \mathcal{M}_{\gamma\gamma} \mathcal{M}_{\gamma} \right]}{\left|\mathcal{M}_{\gamma}\right|} + \mathcal{O}\left(\alpha^4\right). 
\end{equation}
A value of \rtg which remains close to unity for all $Q^2$ and $\epsilon$ indicates minimal hard TPE contribution, while if the \pp cross section exceeds the
\ep cross section ($R_{2\gamma}>1$) by at least several percent for $Q^2\gtrsim2$ (GeV/$c$)$^2$ hard TPE could be responsible for the entirety
of the form factor ratio discrepancy \cite{BerFFPhysRevC.90.015206,schmidt}.

\section{The OLYMPUS Experiment}

\begin{figure}[thb!]
\centerline{\includegraphics[width=0.7\textwidth]{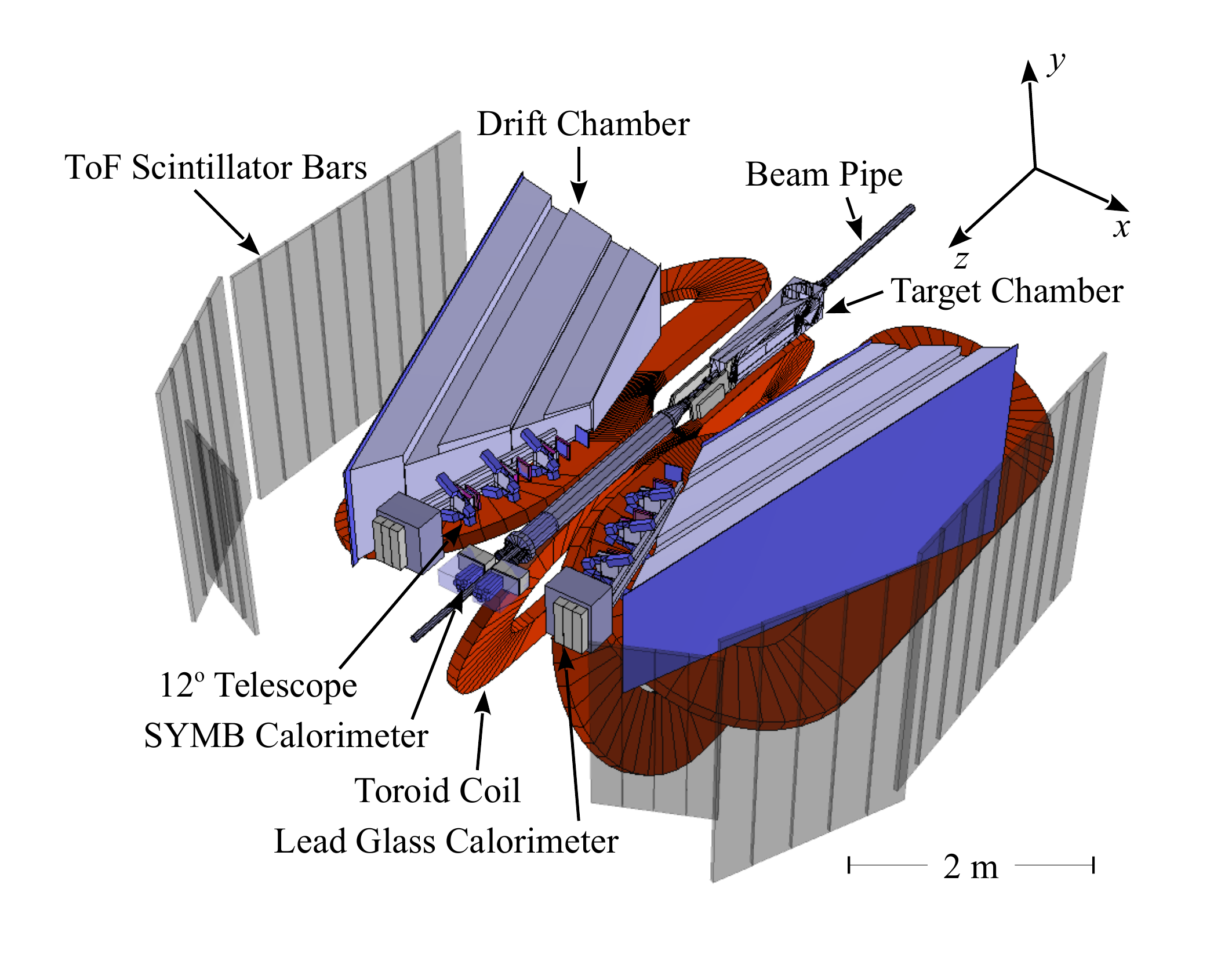}}
\caption{Schematic of the OLYMPUS detector, with the major elements labeled.  The lepton beams passed through the target chamber in the $+\hat{\mathbf{z}}$ direction. }
\label{fig:det}
\end{figure}

The OLYMPUS experiment was designed to measure \rtg by exclusively reconstructing the elastic scattering of 2 GeV electrons and positrons from
a fixed proton target, providing a kinematic reach of $(0.4 \leq \epsilon \leq 0.9)$, $(0.6 \leq Q^2 \leq 2.2)$ (GeV/$c$)$^2$.  The experiment collected
data in 2012 and 2013 at the DORIS storage ring at the Deutsches Elektronen Synchrotron (DESY), Hamburg, Germany.  The stored $e^-$ and $e^+$ beams were incident on a windowless
H$_2$ gas target \cite{Bernauer201420}.  As shown in Figure \ref{fig:det}, the target chamber was installed in a toroidal spectrometer which was adapted
from the detector used at the BLAST experiment at MIT-Bates \cite{Hasell:2009zza}.  Walls of time-of-flight scintillator paddles provided a trigger
signal for the tracking detectors and rough particle identification information, while large-acceptance drift chambers permitted exclusive reconstruction
of leptons and protons trajectories from elastic scattering events.  The lepton beam species was alternated daily
to control long-period systematic effects on the measurement of the \pp and \ep cross sections, and the detector positions and magnetic field were surveyed
in detail to properly account for the acceptance differences for \pp and \ep events \cite{Bernauer20169}.  Using the precise survey of the detector system and magnetic field in conjunction
with a newly written generator for radiative \pmp events \cite{schmidt,russell}, a detailed Geant4 \cite{Agostinelli:2002hh} simulation was developed to account
for the effects of non-hard TPE lepton charge odd effects as well as the detector acceptance and efficiency.  A detailed description of the experiment may
be found in Reference \cite{Milner:2014}.

In addition to reconstructing the elastic \pmp cross sections, determination of \rtg required precise determination of the relative integrated luminosity of electron
and positron scattering data collected by the experiment.  Thus, OLYMPUS employed three independent systems to monitor the relative luminosity:
\begin{enumerate}
\itemsep0em
 \item a ``slow control'' system which combined the recorded beam current with a molecular flow simulation of the target gas density \cite{henderson},
 \item reconstruction of forward ($12^\circ$) elastic scattering events (where \rtg must be close to unity) in dedicated high-resolution tracking telescopes \cite{henderson}, and
 \item a calorimetric measurement of forward lepton-lepton and lepton-proton scattering events at very forward angles ($1.2^\circ$) \cite{PerezBenito20166,SCHMIDT2018112}.
\end{enumerate}
Each of these methods produced consistent results, well within the uncertainty goals of the experiment.  The combination of the latter two methods additionally
allowed extraction of \rtg at $\epsilon\approx 0.98$.  More details on these analyses may be founded in the cited references.

\section{Results}

\begin{figure}[thb!]
    \centering
    \includegraphics[width=0.55\columnwidth]{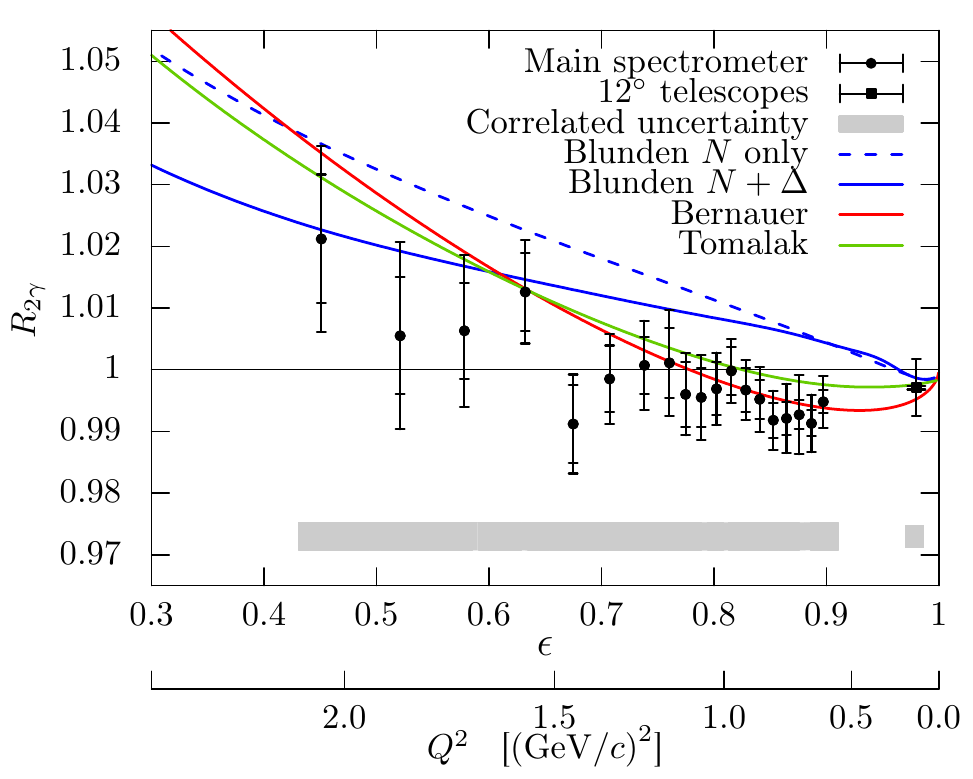}
    \caption{Results of the measurement of \rtg from the OLYMPUS experiment.  The inner error bars represent the statistical uncertainty, the outer error bars the
    total point-to-point uncertainty including systematic contributions, and the gray band the correlated uncertainty due to the relative luminosity extraction.
    Also shown are theoretical \cite{PhysRevC.95.065209,PhysRevD.96.096001} and phenomenological \cite{BerFFPhysRevC.90.015206} predictions.  (Figure adapted from \cite{oPRL}.)}
    \label{fig:money}
\end{figure}

\begin{figure}[thb!]
    \centering
    \includegraphics[width=0.55\columnwidth]{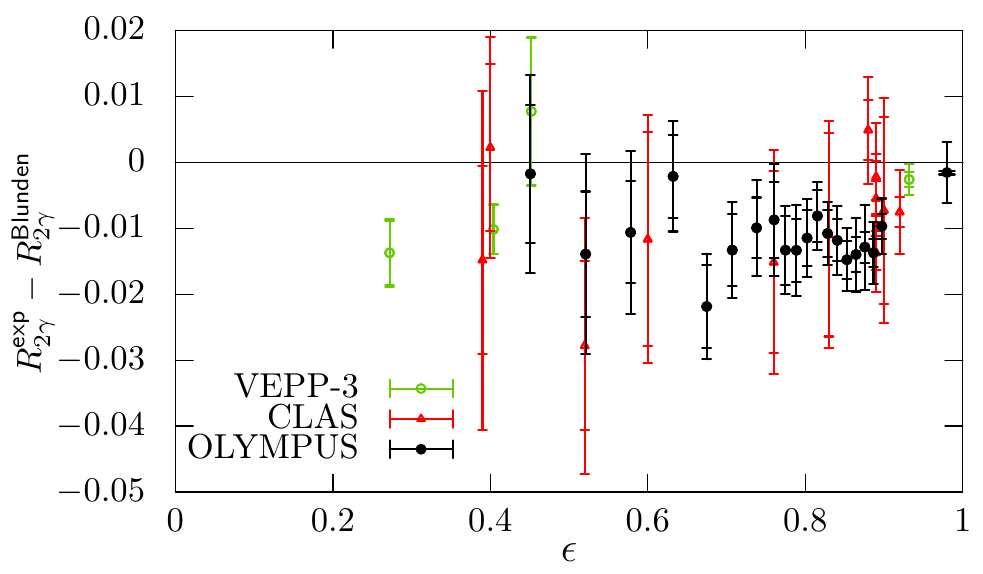}
    \caption{Comparison of the results of the three modern TPE experiments \cite{oPRL,vepp3PhysRevLett.114.062005,PhysRevLett.114.062003,PhysRevC.95.065201}
    to the Blunden dispersive model \cite{PhysRevC.95.065209}, showing that the
    three experiments consistently show a value of \rtg less than the prediction across their respective kinematic ranges. (Figure adapted from \cite{oPRL}.)}
    \label{fig:blund}
\end{figure}

The determination of \rtg by the OLYMPUS experiment is shown in Figure \ref{fig:money} \cite{oPRL}, along with several theoretical and phenomenological
predictions.  The OLYMPUS results show that \rtg remains near unity at small values of $Q^2$ (possibly dropping below one) before increasing
to $\sim$2\% above unity at the high-$Q^2$ end of the OLYMPUS kinematic reach.  While consistent with some phenomenological models, this result implies a
value of \rtg that is somewhat below theoretical models, such as the calculation presented in Reference \cite{PhysRevC.95.065209}, that seek to explain the entire form
factor discrepancy through TPE.  Two other experiments, at the VEPP-3 storage ring in Novosibirsk, Russia and the CLAS experiment at Jefferson Lab,
additionally measured \rtg using different techniques than OLYMPUS and at generally lower values of $Q^2$ \cite{oPRL,vepp3PhysRevLett.114.062005,PhysRevLett.114.062003,PhysRevC.95.065201}.
Figure \ref{fig:blund} shows the combined data of the three experiments compared to the calculation of the Blunden dispersive model corresponding to the kinematics
of each experiment as a means of comparing the results of the experiments in the 2D $(Q^2$,$\epsilon)$ space.
In this comparison, all three experiments consistently show values of \rtg lower than the theoretical prediction.

\section{Discussion}

The OLYMPUS experiment successfully measured \rtg to better than 1\% uncertainty up to $Q^2\approx 2.2$ (GeV/$c$)$^2$.  OLYMPUS and the other
modern TPE experiments measured a relatively modest hard TPE contribution that may be consistent both with resolving the form factor discrepancy and
leaving a significant portion of the problem open.  These results suggest that measurements of the elastic \ep cross section at higher $Q^2$ will be required
to firmly determine the contribution of hard TPE to the form factor discrepancy.  While such experiments will be difficult due to the rapidly decreasing
elastic cross section and increasing non-elastic contributions to \ep scattering with increasing $Q^2$, plans for such experiments are currently under
development \cite{bpos,rpos,axpos}.

\bibliographystyle{JHEP}
\bibliography{references}

\end{document}